\documentclass[manuscript, sigconf]{acmart}
\usepackage[utf8]{inputenc}
\usepackage{amsmath}
\usepackage{algorithm}
\usepackage{algpseudocode}
\usepackage{xcolor}
\definecolor{hgfblue}{RGB}{0, 90, 160}
\colorlet{hgfblue10}{hgfblue!10!white}
\colorlet{hgfblue20}{hgfblue!20!white}
\colorlet{hgfblue30}{hgfblue!30!white}
\colorlet{hgfblue40}{hgfblue!40!white}
\colorlet{hgfblue50}{hgfblue!50!white}
\colorlet{hgfblue60}{hgfblue!60!white}
\colorlet{hgfblue70}{hgfblue!70!white}
\colorlet{hgfblue80}{hgfblue!80!white}
\colorlet{hgfblue90}{hgfblue!90!white}

\definecolor{hgfdarkblue}{RGB}{0, 40, 100}
\colorlet{hgfdarkblue10}{hgfdarkblue!10!white}
\colorlet{hgfdarkblue20}{hgfdarkblue!20!white}
\colorlet{hgfdarkblue30}{hgfdarkblue!30!white}
\colorlet{hgfdarkblue40}{hgfdarkblue!40!white}
\colorlet{hgfdarkblue50}{hgfdarkblue!50!white}
\colorlet{hgfdarkblue60}{hgfdarkblue!60!white}
\colorlet{hgfdarkblue70}{hgfdarkblue!70!white}
\colorlet{hgfdarkblue80}{hgfdarkblue!80!white}
\colorlet{hgfdarkblue90}{hgfdarkblue!90!white}

\definecolor{hgflightblue}{RGB}{20, 200, 255}
\colorlet{hgflightblue10}{hgflightblue!10!white}
\colorlet{hgflightblue20}{hgflightblue!20!white}
\colorlet{hgflightblue30}{hgflightblue!30!white}
\colorlet{hgflightblue40}{hgflightblue!40!white}
\colorlet{hgflightblue50}{hgflightblue!50!white}
\colorlet{hgflightblue60}{hgflightblue!60!white}
\colorlet{hgflightblue70}{hgflightblue!70!white}
\colorlet{hgflightblue80}{hgflightblue!80!white}
\colorlet{hgflightblue90}{hgflightblue!90!white}

\definecolor{hgfgreen}{RGB}{140, 180, 35}
\colorlet{hgfgreen10}{hgfgreen!10!white}
\colorlet{hgfgreen20}{hgfgreen!20!white}
\colorlet{hgfgreen30}{hgfgreen!30!white}
\colorlet{hgfgreen40}{hgfgreen!40!white}
\colorlet{hgfgreen50}{hgfgreen!50!white}
\colorlet{hgfgreen60}{hgfgreen!60!white}
\colorlet{hgfgreen70}{hgfgreen!70!white}
\colorlet{hgfgreen80}{hgfgreen!80!white}
\colorlet{hgfgreen90}{hgfgreen!90!white}

\definecolor{hgfgray}{RGB}{90, 105, 110}
\colorlet{hgfgray10}{hgfgray!10!white}
\colorlet{hgfgray20}{hgfgray!20!white}
\colorlet{hgfgray30}{hgfgray!30!white}
\colorlet{hgfgray40}{hgfgray!40!white}
\colorlet{hgfgray50}{hgfgray!50!white}
\colorlet{hgfgray60}{hgfgray!60!white}
\colorlet{hgfgray70}{hgfgray!70!white}
\colorlet{hgfgray80}{hgfgray!80!white}
\colorlet{hgfgray90}{hgfgray!90!white}

\definecolor{hgfhighlight}{RGB}{205, 238, 251}
\colorlet{hgfhighlight10}{hgfhighlight!10!white}
\colorlet{hgfhighlight20}{hgfhighlight!20!white}
\colorlet{hgfhighlight30}{hgfhighlight!30!white}
\colorlet{hgfhighlight40}{hgfhighlight!40!white}
\colorlet{hgfhighlight50}{hgfhighlight!50!white}
\colorlet{hgfhighlight60}{hgfhighlight!60!white}
\colorlet{hgfhighlight70}{hgfhighlight!70!white}
\colorlet{hgfhighlight80}{hgfhighlight!80!white}
\colorlet{hgfhighlight90}{hgfhighlight!90!white}

\definecolor{hgfmint}{RGB}{5, 229, 186}
\colorlet{hgfmint10}{hgfmint!10!white}
\colorlet{hgfmint20}{hgfmint!20!white}
\colorlet{hgfmint30}{hgfmint!30!white}
\colorlet{hgfmint40}{hgfmint!40!white}
\colorlet{hgfmint50}{hgfmint!50!white}
\colorlet{hgfmint60}{hgfmint!60!white}
\colorlet{hgfmint70}{hgfmint!70!white}
\colorlet{hgfmint80}{hgfmint!80!white}
\colorlet{hgfmint90}{hgfmint!90!white}

\definecolor{hgfpale}{RGB}{236, 251, 253}
\colorlet{hgfpale10}{hgfpale!10!white}
\colorlet{hgfpale20}{hgfpale!20!white}
\colorlet{hgfpale30}{hgfpale!30!white}
\colorlet{hgfpale40}{hgfpale!40!white}
\colorlet{hgfpale50}{hgfpale!50!white}
\colorlet{hgfpale60}{hgfpale!60!white}
\colorlet{hgfpale70}{hgfpale!70!white}
\colorlet{hgfpale80}{hgfpale!80!white}
\colorlet{hgfpale90}{hgfpale!90!white}

\definecolor{hgfaerospace}{RGB}{80, 200, 170}
\definecolor{hgfast}{named}{hgfaerospace}

\colorlet{hgfaerospace10}{hgfaerospace!10!white}
\colorlet{hgfaerospace20}{hgfaerospace!20!white}
\colorlet{hgfaerospace30}{hgfaerospace!30!white}
\colorlet{hgfaerospace40}{hgfaerospace!40!white}
\colorlet{hgfaerospace50}{hgfaerospace!50!white}
\colorlet{hgfaerospace60}{hgfaerospace!60!white}
\colorlet{hgfaerospace70}{hgfaerospace!70!white}
\colorlet{hgfaerospace80}{hgfaerospace!80!white}
\colorlet{hgfaerospace90}{hgfaerospace!90!white}

\colorlet{hgfast10}{hgfast!10!white}
\colorlet{hgfast20}{hgfast!20!white}
\colorlet{hgfast30}{hgfast!30!white}
\colorlet{hgfast40}{hgfast!40!white}
\colorlet{hgfast50}{hgfast!50!white}
\colorlet{hgfast60}{hgfast!60!white}
\colorlet{hgfast70}{hgfast!70!white}
\colorlet{hgfast80}{hgfast!80!white}
\colorlet{hgfast90}{hgfast!90!white}

\definecolor{hgfearthandenvironment}{RGB}{50, 100, 105}
\definecolor{hgfee}{named}{hgfearthandenvironment}

\colorlet{hgfearthandenvironment10}{hgfearthandenvironment!10!white}
\colorlet{hgfearthandenvironment20}{hgfearthandenvironment!20!white}
\colorlet{hgfearthandenvironment30}{hgfearthandenvironment!30!white}
\colorlet{hgfearthandenvironment40}{hgfearthandenvironment!40!white}
\colorlet{hgfearthandenvironment50}{hgfearthandenvironment!50!white}
\colorlet{hgfearthandenvironment60}{hgfearthandenvironment!60!white}
\colorlet{hgfearthandenvironment70}{hgfearthandenvironment!70!white}
\colorlet{hgfearthandenvironment80}{hgfearthandenvironment!80!white}
\colorlet{hgfearthandenvironment90}{hgfearthandenvironment!90!white}

\colorlet{hgfee10}{hgfee!10!white}
\colorlet{hgfee20}{hgfee!20!white}
\colorlet{hgfee30}{hgfee!30!white}
\colorlet{hgfee40}{hgfee!40!white}
\colorlet{hgfee50}{hgfee!50!white}
\colorlet{hgfee60}{hgfee!60!white}
\colorlet{hgfee70}{hgfee!70!white}
\colorlet{hgfee80}{hgfee!80!white}
\colorlet{hgfee90}{hgfee!90!white}

\definecolor{hgfenergy}{RGB}{255, 210, 40}

\colorlet{hgfenergy10}{hgfenergy!10!white}
\colorlet{hgfenergy20}{hgfenergy!20!white}
\colorlet{hgfenergy30}{hgfenergy!30!white}
\colorlet{hgfenergy40}{hgfenergy!40!white}
\colorlet{hgfenergy50}{hgfenergy!50!white}
\colorlet{hgfenergy60}{hgfenergy!60!white}
\colorlet{hgfenergy70}{hgfenergy!70!white}
\colorlet{hgfenergy80}{hgfenergy!80!white}
\colorlet{hgfenergy90}{hgfenergy!90!white}

\definecolor{hgfhealth}{RGB}{210, 50, 100}

\colorlet{hgfhealth10}{hgfhealth!10!white}
\colorlet{hgfhealth20}{hgfhealth!20!white}
\colorlet{hgfhealth30}{hgfhealth!30!white}
\colorlet{hgfhealth40}{hgfhealth!40!white}
\colorlet{hgfhealth50}{hgfhealth!50!white}
\colorlet{hgfhealth60}{hgfhealth!60!white}
\colorlet{hgfhealth70}{hgfhealth!70!white}
\colorlet{hgfhealth80}{hgfhealth!80!white}
\colorlet{hgfhealth90}{hgfhealth!90!white}

\definecolor{hgfinformation}{RGB}{160, 35, 90}
\definecolor{hginfo}{named}{hgfinformation}

\colorlet{hgfinformation10}{hgfinformation!10!white}
\colorlet{hgfinformation20}{hgfinformation!20!white}
\colorlet{hgfinformation30}{hgfinformation!30!white}
\colorlet{hgfinformation40}{hgfinformation!40!white}
\colorlet{hgfinformation50}{hgfinformation!50!white}
\colorlet{hgfinformation60}{hgfinformation!60!white}
\colorlet{hgfinformation70}{hgfinformation!70!white}
\colorlet{hgfinformation80}{hgfinformation!80!white}
\colorlet{hgfinformation90}{hgfinformation!90!white}

\colorlet{hgfinfo10}{hgfinformation!10!white}
\colorlet{hgfinfo20}{hgfinformation!20!white}
\colorlet{hgfinfo30}{hgfinformation!30!white}
\colorlet{hgfinfo40}{hgfinformation!40!white}
\colorlet{hgfinfo50}{hgfinformation!50!white}
\colorlet{hgfinfo60}{hgfinformation!60!white}
\colorlet{hgfinfo70}{hgfinformation!70!white}
\colorlet{hgfinfo80}{hgfinformation!80!white}
\colorlet{hgfinfo90}{hgfinformation!90!white}

\definecolor{hgfmatter}{RGB}{240, 120, 30}

\colorlet{hgfmatter10}{hgfmatter!10!white}
\colorlet{hgfmatter20}{hgfmatter!20!white}
\colorlet{hgfmatter30}{hgfmatter!30!white}
\colorlet{hgfmatter40}{hgfmatter!40!white}
\colorlet{hgfmatter50}{hgfmatter!50!white}
\colorlet{hgfmatter60}{hgfmatter!60!white}
\colorlet{hgfmatter70}{hgfmatter!70!white}
\colorlet{hgfmatter80}{hgfmatter!80!white}
\colorlet{hgfmatter90}{hgfmatter!90!white}

\usepackage{graphicx}
\usepackage{subcaption}
\usepackage{tabularx}
\usepackage{tikz}
\usepackage{cleveref}
\usepackage{float}
\usepackage{hyperref}
\usepackage{todonotes}
\usetikzlibrary{
  arrows.meta,
  positioning,
  shadows,
  shapes.multipart,
  calc,
  backgrounds,
  decorations.pathreplacing,
  fit
}
\AtBeginDocument{%
  }

\setcopyright{none}
\renewcommand\footnotetextcopyrightpermission[1]{}

\begin{document}

\title{Theoretical and Empirical Analysis of Sparse Convolution}


\author{Keshvi Tuteja}
\affiliation{%
  \institution{Scientific Computing Center, Karlsruhe Institute of Technology}
  \country{Germany}
}
\email{keshvi.tuteja@kit.edu}

\author{Yu-Hsiang Tsai}
\affiliation{%
  \institution{Technical University of Munich}
  \country{Germany}
}
\email{yu-hsiang.tsai@tum.de}

\author{Gregor Olenik}
\affiliation{%
  \institution{Technical University of Munich}
  \country{Germany}
}
\email{gregor.olenik@tum.de}

\author{Markus G{\"o}tz}
\affiliation{%
  \institution{Helmholtz AI and Scientific Computing Center, Karlsruhe Institute of Technology}
  \country{Germany}
}
\email{markus.goetz@kit.edu}

\author{Achim Streit}
\affiliation{%
  \institution{Scientific Computing Center, Karlsruhe Institute of Technology}
  \country{Germany}
}
\email{achim.streit@kit.edu}

\author{Hartwig Anzt}
\affiliation{%
  \institution{
  Chair of Computational Mathematics, Technical University of Munich,
  University of Tennessee}
  \country{Germany, USA}
}
\email{hartwig.anzt@tum.de}

\author{Charlotte Debus}
\affiliation{%
  \institution{Scientific Computing Center, Karlsruhe Institute of Technology}
  \country{Germany}
}
\email{charlotte.debus@kit.edu}

\renewcommand{\shortauthors}{Tuteja et al.}

\begin{abstract}
Deep neural networks for image processing are increasingly deployed on resource-constrained edge devices such as drones, medical systems, and smartphones, where limited computational resources and memory bandwidth pose significant challenges for efficient inference. Pruning during training often produces highly sparse convolutional kernels, offering the potential to significantly reduce computation and memory requirements. However, exploiting this sparsity efficiently at runtime remains challenging due to irregular memory access patterns and workload imbalance across processing elements. In this work, we investigate the standard convolution with sparse weights and analyze its performance characteristics through roofline analysis, showing that a $3\times3$ sparse convolution achieves an arithmetic intensity of approximately 10~FLOP/byte compared to $\sim$96~FLOP/byte for its dense counterpart. We propose techniques for an efficient sparse convolution at high sparsities. Preliminary results demonstrate that our approach achieves upto $1.65\times$ speedup as compared to dense convolution on NVIDIA A100 and upto $1.52\times$ on NVIDIA Jetson Nano (Tegra X1) , motivating further exploration of direct convolution as a practical path to efficient sparse inference on modern accelerators/edge devices.
\end{abstract}




\maketitle

\section{Introduction}
\label{sec:intro}
Convolutional neural networks (CNNs)~\cite{oshea2015introductionconvolutionalneuralnetworks} have become foundational to a wide range of applications, including medical imaging, autonomous driving, robotics, surveillance, and mobile vision systems. The growing demand for real time, on device intelligence has shifted many CNN workloads from data centers to edge platforms, where processing data locally reduces latency, communication costs, and privacy concerns. However, edge devices such as embedded GPUs, FPGAs, and custom ASICs provide only limited memory capacity, power budgets and memory bandwidth~\cite{9708948}. This makes computational efficiency and low memory consumption essential.
In many image processing tasks on edge deployment scenarios, model weights reside permanently in device memory while input feature maps, for example in the form of input images, are streamed at runtime. Due to the limited memory capacity of edge devices, memory footprint reduction by storing only non-zero model weights explicitly via sparse matrix formats becomes an attractive approach for reducing memory traffic and improving inference efficiency.

One of the most promising strategies for reducing the memory footprint and computational cost of CNNs on resource-constrained hardware is weight sparsification~\cite{raihan2020sparse,stamenovic2021weightblockunitexploring}. By pruning a large fraction of the learned weights to zero, sparse networks can substantially reduce both floating-point operations (FLOPs) and memory traffic during inference~\cite{liu2024efficientexpertpruningsparse,hoefler2021sparsitydeeplearningpruning}. Consequently, recent research has focused on pruning networks to extreme sparsity levels, typically between $90\%$ and $99\%$~\cite{kuznedelev2023accurateneuralnetworkpruning}. However, translating these theoretical savings into practical performance gains remains challenging because sparse implementations incur additional execution overhead. In particular, irregular sparsity patterns require indirect accesses to locate non-zero weights and produce uneven workloads across processing elements, resulting in load imbalance and reduced memory efficiency~\cite{gale2020sparsegpukernelsdeep,10.1145/1654059.1654078}.


For a sparse convolution (sparse weights, dense images), We identify two principal obstacles to high performance: (1)~\emph{inefficient sparse weight access}, whereby non-contiguous weight layouts generate memory transactions that underutilize memory bandwidth, and (2)~\emph{workload imbalance}, arising because different output channels or spatial tiles encounter varying numbers of non-zero weights, causing some processing elements to idle while others remain saturated. We propose a packed weight representation that addresses the first challenge by enabling coalesced, single-transaction access and improved cache utilization, together with a thread mapping strategy that helps equalize load across threads for the second challenge.

This paper makes three contributions: 
\begin{itemize}
    \item A roofline analysis showing that direct sparse convolution at practical sparsity levels operates at or near the memory-bandwidth bound regime of modern accelerators, motivating our kernel design.
    \item A sparse convolution kernel utilizing compact 32-bit packed weight format combined with a shared-memory tiling for better memory bandwidth utilization.
    \item An evaluation on both an NVIDIA A100 and an edge device: Jetson Nano (Tegra X1) showing consistent speedups over PyTorch's dense convolution baseline, reaching up to $1.65\times$ on the A100 and upto $1.52\times$ on NVIDIA Tegra X1.
\end{itemize}


\section{Related Work}
\label{sec:background}

Convolution kernels can broadly be categorized into three algorithmic families, each with distinct performance trade-offs. The first family relies on \emph{Im2col-based transformations}~\cite{Chellapilla2006HighPerformanceCNN}, which reshape and rearrange the input tensor into a matrix so that convolution reduces to a general matrix–matrix multiply (GEMM)~\cite{Ramirez2024} or a sparse matrix–vector product (SpMV)/ sparse matrix–matrix product (SpMM)~\cite{Elsen_2020_CVPR}. There have also been efforts in improving sparse convolution by improving sparse operations (SpMV/SpMM)~\cite{Tuteja_2025}. While this approach maps well onto highly tuned BLAS libraries, it typically replicates input data, inflating memory consumption by a factor of $\approx K^2$, where $K$ is the kernel size. Therefore a $3\times3$ convolution would inflate memory consumption by $9\times$ and a $5\times5$ by $25\times$ and so on. These approaches also fail to fully exploit the data reuse hierarchy available in modern GPU caches and scratchpad memories. While there has been research in the direction of better memory usage in such convolutions ~\cite{Lu_2022}, it remains limited to dense convolution~\cite{zhou2021characterizingdemystifyingimplicitconvolution}.~\cite{park2017fastercnnsdirectsparse} also discusses and implements a sparse convolution which is based on sparse matrix-dense matrix multiplication but this only supports execution on CPUs. 

The second family employs \emph{transformation based convolution} such as the Fast Fourier Transform (FFT) and the Winograd algorithm. FFT-based convolution is suitable for large filters but suffers significant overhead for the small kernel sizes (e.g., $3\times3$, $5\times5$, the typical kernel sizes used in deep learning). Therefore, it is mainly used for signal processing tasks rather than deep learning. Another approach to compute convolution is through Winograd minimal filtering algorithm~\cite{alam2022winogradconvolutiondeepneural}. Dense $3\times3$ convolutions are typically accelerated using Winograd algorithms, which reduce the number of floating-point multiplications through small domain transforms~\cite{Zlateski2018FFTWinograd}. The widely used $F(2\times2,3\times3)$ Winograd algorithm decreases the multiplication count by approximately $2.25\times$ compared to direct convolution~\cite{10.1145/3524069}, making it the preferred implementation for small dense convolutions in libraries such as cuDNN. However, Winograd relies on dense transform matrices and dense element-wise multiplications in the transformed domain. When applied to highly sparse convolution kernels, these transformations destroy the original sparsity pattern and introduce additional arithmetic and memory overhead, substantially reducing the potential benefits of weight pruning. Consequently, the next family: direct sparse convolution remains a more suitable approach for exploiting high levels of unstructured sparsity. 

The third family performs \emph{direct convolution}, sliding the kernel across the input without any data transformation. Direct convolution preserves the full locality of the computation and avoids the overhead of format conversion, making it a natural fit for sparse convolution kernels. 

\section{Theoretical Background}
\label{sec:bg}
Standard multi-channel image convolution is represented as:
\begin{equation}
\begin{split}
C_{h,w,oc} = \sum_{ic=0}^{IC-1} \hspace{2mm}\sum_{k_h=\max(0,P_h-h)}^{\min(K_h-1,H+P_h-h-1)} 
\hspace{2mm} \sum_{k_w=\max(0,P_w-w)}^{\min(K_w-1,W+P_w-w-1)} \\ X_{h+k_h-P_h,\,w+k_w-P_w,\,ic} \cdot \mathcal{W}_{k_h,k_w,ic,oc}
\end{split}
\label{eq:dense_conv}
\end{equation}

\begin{equation}
H_{out} = \left\lfloor \frac{H + 2P_h - K_h}{S_h} \right\rfloor + 1,
\qquad
W_{out} = \left\lfloor \frac{W + 2P_w - K_w}{S_w} \right\rfloor + 1
\label{eq:conv_output_size}
\end{equation}

where:
\begin{itemize}
    \item $C$ is the output image with multiple channels, defined on $h \in [0,H_{out})$, $w \in [0,W_{out})$,
    \item $S_h, S_w$ denote the stride in each spatial dimension (set to $1$ for the unstrided case used above),
    \item $ic$ and $oc$ denote the number of input and output channels,
    \item $K_h, K_w$ denote the kernel height and width,
    \item $P_h, P_w$ denote the (symmetric) zero-padding applied to $X$ in each spatial dimension,
    \item $X$ is the input image with multiple channels, defined only on $h \in [0,H)$, $w \in [0,W)$,
    \item $\mathcal{W}$ are the convolution weights,

    \item $h$ and $w$ are the output coordinates, $h \in [0,H_{out})$, $w \in [0,W_{out})$.
\end{itemize}

In \cref{eq:dense_conv}, one output element $C_{h,w,oc}$ is computed by accumulating
contributions from all input channels and all kernel positions, restricted to the
valid summation range so that $X$ is never indexed outside its defined domain
$[0,H)\times[0,W)$. Unlike dense convolution, sparse convolution only iterates
over nonzero weights, thereby reducing memory traffic and arithmetic operations.
Sparse convolution can therefore be written as:
\begin{equation}
C_{h,w,oc}
=
\sum_{j \in \mathcal{N}(oc,h,w)}
v_j \cdot
X_{\Delta_h^{(j)},\,\Delta_w^{(j)},\,ic^{(j)}}
\label{eq:sparse_conv}
\end{equation}
where $\Delta_h^{(j)} = h+p_y^{(j)}-P_h$ and $\Delta_w^{(j)} = w+p_x^{(j)}-P_w$
denote the padding-shifted input coordinates addressed by nonzero weight $j$,
and:
\begin{itemize}
    \item $\mathcal{N}(oc,h,w) = \{\, j \in \mathcal{N}(oc) : 0 \le \Delta_h^{(j)} < H,\ 0 \le \Delta_w^{(j)} < W \,\}$
          restricts $oc$'s nonzero weights to those whose input position lies
          within the valid image bounds at this output location,
    \item $\mathcal{N}(oc)$ denotes the full set of nonzero weights associated
          with output channel $oc$,
    \item $v_j$ is the value of the $j$-th nonzero weight,
    \item $X$ is the input image with multiple channels, defined only on $h \in [0,H)$, $w \in [0,W)$,
    \item $ic^{(j)}$ is the corresponding input channel,
    \item $(p_x^{(j)}, p_y^{(j)})$ are the kernel offsets of the $j$-th nonzero weight,
    \item $P_h, P_w$ denote the symmetric zero-padding applied to $X$, matching
          Eq.~\ref{eq:dense_conv}.
\end{itemize}

\begin{figure*}[t]
    \centering
\scalebox{1.1}{
    \input{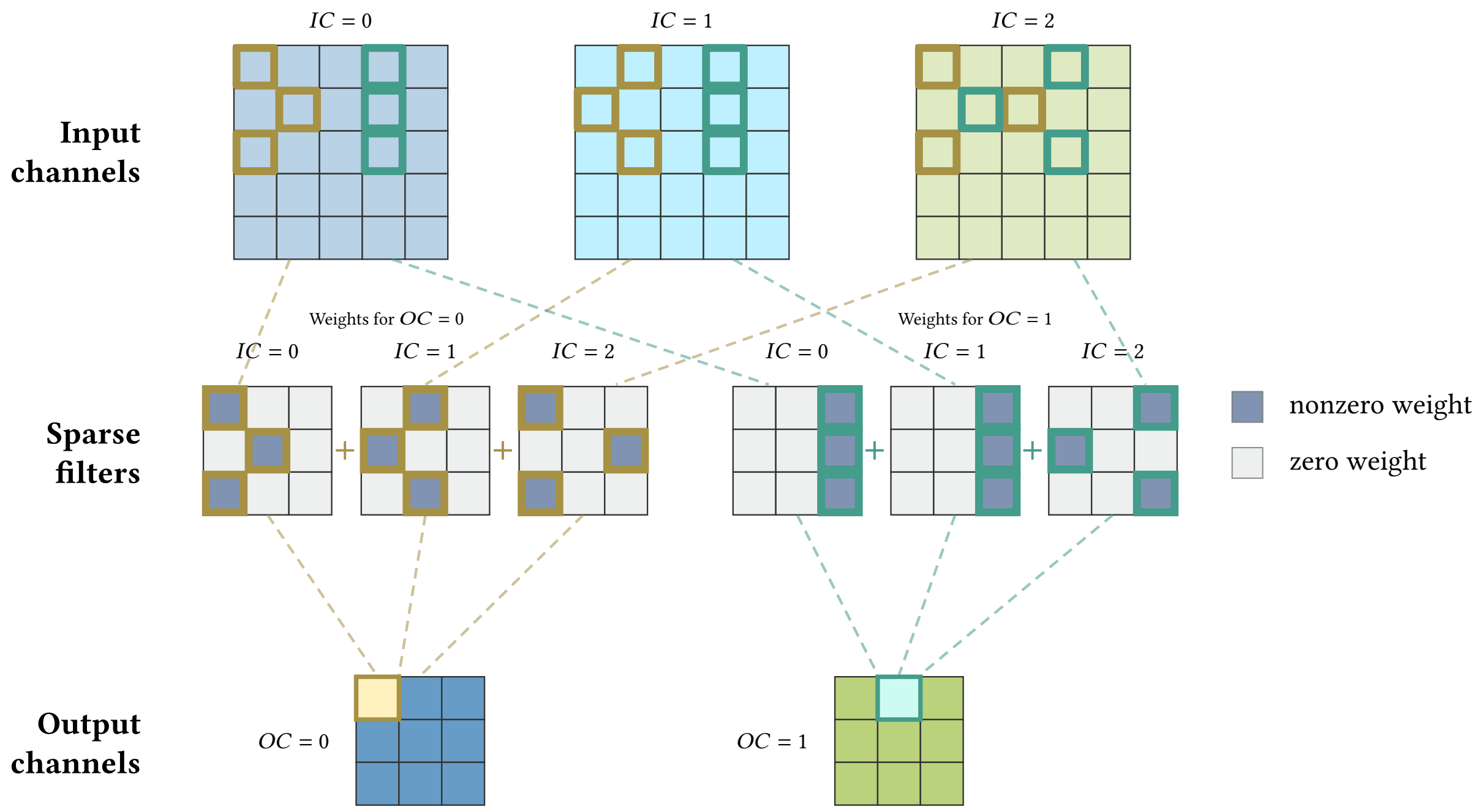}
}
    \caption{Sparse convolution with sparse weights. Here, the input image has 3 input channels (IC), output image has 2 output channels (OC). OC denotes the number of output channels, with each output channel generated using a distinct set of convolutional kernels spanning all input channels. The convolution filter has a 3x3 weight matrix for each OC and IC. }
    \label{fig:spconv}
\end{figure*}

The sparse convolution is visually represented in \cref{fig:spconv}. 

\section{Roofline Modelling}
\label{sec:roofline}

To understand the performance characteristics of sparse and dense convolution,
we apply roofline modeling~\cite{10.1145/1498765.1498785} to characterize whether
each formulation is compute-bound or memory-bound on the target hardware. The resulting analysis guides the design of our sparse convolution kernel by identifying the dominant performance bottleneck.  \\
A key metric in roofline modelling is Arithmetic intensity (AI). AI is defined as the ratio of floating-point operations done per byte of memory accessed:
\begin{equation}
\label{eq:ai}
    \text{AI} = \frac{Computation[\text{FLOPs}]}{Communicaton[\text{Bytes}]}.
\end{equation}

For example, for the NVIDIA A100 GPU with a peak floating-point throughput
of $19.5$~TFLOP/s and a memory bandwidth of $2$~TB/s~\cite{nvidia_a100_datasheet},
the FP32 ridge point is approximately $9.75$~FLOP/byte. Kernels with arithmetic intensity below this threshold are memory-bound, and those above are compute-bound. Edge accelerators shift this ridge point towards higher values: the Jetson AGX Orin (64GB), with $5.32$~TFLOP/s FP32
and $204.8$~GB/s of bandwidth, has a ridge point of approximately $25.98$~FLOP/byte which is more than $2.5\times$ that of the A100, while the Jetson Orin Nano (8GB), at $5.0$~TFLOP/s and only $68$~GB/s, reaches approximately $73.53$~FLOP/byte, nearly an order of magnitude higher than the A100. In contrast, the original Jetson Nano (Tegra X1), at $0.236$~TFLOP/s and $25.6$~GB/s, lands at approximately $9.22$~FLOP/byte: almost identical to the A100's ridge point despite roughly $80\times$ lower absolute throughput. This demonstrates that the ridge point is governed by the ratio of peak compute performance to memory bandwidth rather than their absolute values. Consequently, a kernel with fixed arithmetic intensity may be compute-bound on one architecture but memory-bound on another, even if the latter offers higher peak performance. Since edge devices typically have higher ridge points, more kernels fall into the memory-bound regime, making bandwidth optimization increasingly important.

From \cref{eq:ai}, we calculate the arithmetic intensity by estimating both the communication and the computational cost of dense and sparse convolutions. Consider a convolution with batch size $N$, input channels $IC$, output channels $OC$, kernel size $K\times K$, and output spatial dimensions $H_{\mathrm{out}}\times W_{\mathrm{out}}$. Let $P$ denote the size of each floating-point value in bytes.

\paragraph{Dense convolution.}
The memory traffic/communication cost consists of reading the input image/s, reading the weight tensor, and writing the output image/s:
\begin{equation}
\begin{aligned}
\mathrm{Mem}_{\mathrm{dense}} =
\Big(
&NHW\cdot IC
+K^2IC\,OC \\
&+H_{\mathrm{out}}W_{\mathrm{out}}N\,OC
\Big)P 
\end{aligned}
\label{eq:mem_dense}
\end{equation}

The corresponding computational cost is

\begin{equation}
\mathrm{FLOPs}_{\mathrm{dense}}
=
2\,N\,H_{\mathrm{out}}W_{\mathrm{out}}\,OC\,K^2IC,
\label{eq:flops_dense}
\end{equation}

where the factor of two accounts for one multiply and one addition per multiply accumulate operation.

\paragraph{Sparse convolution.}
Assume an unstructured weight sparsity of $S\in[0,1)$, such that only a fraction $(1-S)$ of the weights are retained. Since the input and output activations remain dense, only the weight tensor is compressed. The resulting memory traffic becomes

\begin{equation}
\begin{aligned}
\mathrm{Mem}_{\mathrm{sparse}}
=
&
\left(
NHW\cdot IC
+
NH_{\mathrm{out}}W_{\mathrm{out}}\,OC
\right)P
\\
&
+
(1-S)K^2IC\,OC
\left(
P+P_{\mathrm{idx}}
\right),
\end{aligned}
\label{eq:mem_sparse}
\end{equation}

where $P_{\mathrm{idx}}$ denotes the storage required for the index metadata associated with each nonzero weight. In our implementation, each nonzero weight is stored as a 16-bit FP16 value together with 16 bits of indexing information in a single 32-bit packed entry, requiring 4 bytes per nonzero ($P=2$ bytes and $P_{\mathrm{idx}}=2$ bytes). During convolution, the FP16 weights are converted to FP32 and accumulated in FP32 registers without additional per-nonzero accumulator storage.

The computational cost decreases proportionally with the number of retained weights:

\begin{equation}
\mathrm{FLOPs}_{\mathrm{sparse}}
=
2\,N\,H_{\mathrm{out}}W_{\mathrm{out}}\,OC\,
(1-S)\,
K^2IC.
\label{eq:flops_sparse}
\end{equation}

\subsection{Impact of Sparsity}

\begin{figure*}[!h]
    \centering
    \begin{subfigure}{0.48\textwidth}
        \includegraphics[width=\textwidth]{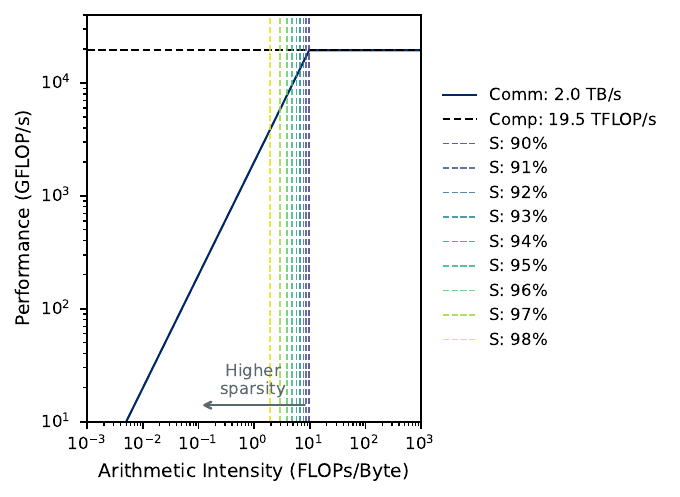}
        \caption{NVIDIA A100}
        \label{fig:a100_heir}
    \end{subfigure}
    \begin{subfigure}{0.48\textwidth}
        \includegraphics[width=\textwidth]{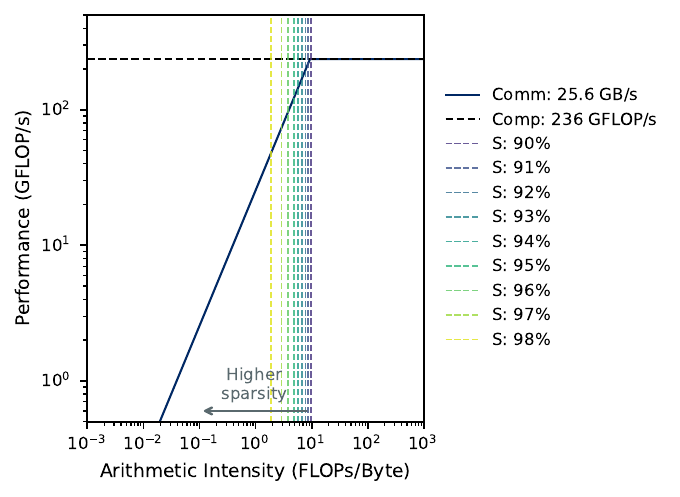}
        \caption{Jetson Nano}
        \label{fig:tegra_heir}
    \end{subfigure}
    \caption{Roofline model for sparse convolution on NVIDIA A100 and Jetson Nano (NVIDIA Tegra X1)}
    \label{fig:bar}
\end{figure*}

Using \cref{eq:mem_sparse} and \cref{eq:flops_sparse}, the arithmetic intensities for direct convolution across varying sparsities were computed. The values of arithmetic intensities for sparsities $S \in [0.90,\,0.98]$ with increments of $0.01$ are reported in \cref{tab:ai_kernel}. A fixed configuration of $H = W = 256$, $IC = 32$, $OC = 64$, $N = 32$ and $K=3$ was used. The corresponding roofline plots for NVIDIA A100 and NVIDIA Jetson Nano (Tegra X1) are presented in figures \ref{fig:a100_heir} and \ref{fig:tegra_heir} respectively.


\setlength{\tabcolsep}{3pt}

\begin{table}[t]
\centering
\caption{Arithmetic intensity (FLOP/byte) for varying weight sparsities.}
\label{tab:ai_kernel}
\begin{tabular}{lccccccccc}
\toprule
Sparsity & 90 & 91 & 92 & 93 & 94 & 95 & 96 & 97 & 98 \\
\midrule
AI & 9.6 & 8.6 & 7.7 & 6.7 & 5.7 & 4.8 & 3.8 & 2.9 & 1.9 \\
\bottomrule
\end{tabular}
\end{table}

\setlength{\tabcolsep}{6pt} 

The corresponding arithmetic intensity for dense convolution is around $96$ (computed using \cref{eq:mem_dense} and \cref{eq:flops_dense}), which is about an order of magnitude larger than the arithmetic intensity values for the sparse convolution. Dense convolution is strongly compute-bound with AI exceeding the ridge point at all the sparsities in \cref{tab:ai_kernel}. Sparse convolution at 90\% sparsity reduces AI by approximately $10\times$, which is consistent with the $(1{-}S)$ reduction in FLOPs. The $S=0.90$ sparse case with an AI of $\sim$9.6~FLOP/byte, falls right at the ridge point of our target GPU. This means that even the smallest inefficiencies in memory access patterns such as uncoalesced loads caused by irregular sparsity structure, will push the kernel below the ridge point and into the memory-bandwidth bound regime. The rest of the values fall below the ridge point, making memory-accesses the main bottleneck for the sparse convolution.



\subsection{Roofline Comparison: Direct vs.\ Winograd}

Since Winograd convolution is the standard high-performance baseline for small kernel sizes and is used by major deep learning frameworks, we also compare sparse direct convolution against Winograd based dense convolution. As already discussed earlier, Winograd based algorithms can reduce the number of operations up to and higher than roughly $2x$. Table~\ref{tab:speedup_a100} and ~\ref{tab:speedup_nano} report the theoretical time based speedup of sparse direct convolution over Winograd dense convolution for varying input and output channel counts for NVIDIA A100 and NVIDIA Jetson nano (Tegra X1) respectively. The analysis was done for $K{=}3$, $H{=}W{=}256$, $N{=}32$, and $S{=}0.9$ and assuming a reduction in operations by $1.5x$. This analysis was done by calculating the execution time for each of those kernels using Roofline analysis for NVIDIA A100. Here, we assumed, execution time to be $max(T_{comp}, T_{comm})$.

\begin{table}[h]
\centering
\caption{Theoretical speedup of sparse direct convolution over Winograd dense
convolution on NVIDIA A100. $K{=}3$, $H{=}W{=}256$, $N{=}32$, $S{=}0.9$.}
\label{tab:speedup_a100}
\begin{tabular}{l rrrrr}
\toprule
 & \multicolumn{5}{c}{Output Channels (OC)} \\
\cmidrule(lr){2-6}
IC & 32 & 64 & 128 & 256 & 512 \\
\midrule
1   & 1.0$\times$ & 1.0$\times$ & 1.0$\times$ & 1.0$\times$ & 1.0$\times$ \\
16  & 1.8$\times$ & 2.2$\times$ & 2.5$\times$ & 2.6$\times$ & 2.7$\times$ \\
32  & 2.8$\times$ & 3.7$\times$ & 4.4$\times$ & 4.9$\times$ & 5.0$\times$ \\
64  & 3.7$\times$ & 5.0$\times$ & 5.0$\times$ & 5.0$\times$ & 5.0$\times$ \\
128 & 4.4$\times$ & 5.0$\times$ & 5.0$\times$ & 5.0$\times$ & 5.0$\times$ \\
\bottomrule
\end{tabular}
\end{table}

\begin{table}[h]
\centering
\caption{Theoretical speedup of sparse direct convolution over Winograd dense
convolution on Jetson Nano (Tegra X1). $K{=}3$, $H{=}W{=}128$, $N{=}8$, $S{=}0.9$.}
\label{tab:speedup_nano}
\begin{tabular}{l rrrrr}
\toprule
 & \multicolumn{5}{c}{Output Channels (OC)} \\
\cmidrule(lr){2-6}
IC & 16 & 24 & 32 & 40 & 48 \\
\midrule
1   & 1.0$\times$ & 1.0$\times$ & 1.0$\times$ & 1.0$\times$ & 1.0$\times$ \\
8  & 1.3$\times$ & 1.5$\times$ & 1.6$\times$ & 1.6$\times$ & 1.7$\times$ \\
16  & 2.0$\times$ & 2.3$\times$ & 2.6$\times$ & 2.8$\times$ & 2.9$\times$ \\
24  & 2.3$\times$ & 3.0$\times$ & 3.3$\times$ & 3.7$\times$ & 4.0$\times$ \\
32 & 2.6$\times$ & 3.3$\times$ & 4.0$\times$ & 4.3$\times$ & 4.7$\times$ \\
\bottomrule
\end{tabular}
\end{table}

At $IC{=}1$, no speedup is achieved because both formulations are memory bound and the execution time is dominated by loading the input and output feature maps rather than the weights at large batch sizes. As $IC$ increases, the weight tensor grows and the sparsity savings become significant, yielding a consistent theoretical speedup of approximately $4.5$--$4.8\times$ over Winograd for channel counts representative of practical CNN layers. 
  
These results confirm that sparse direct convolution is a theoretically compelling alternative to Winograd for high-sparsity regimes, provided that the memory access irregularity introduced by sparse weight formats can be adequately managed. But for this analysis, we take the optimistic estimate for speedup. It is calculated assuming that the communication and computation completely overlap with each other. 

According to prior studies, profiling of cuSPARSE's csrmv (compressed sparse row matrix-vector multiplication) kernel on an NVIDIA A100 shows memory bandwidth utilization of only $2.9$–$11.34\%$ for irregular application matrices, versus $67.79$–$73.67\%$ for dense GEMV (General Matrix-Vector multiplication) on the same hardware, illustrating the gap that pure FLOPs/bytes roofline accounting doesn't fully explain~\cite{liu2024gpu}.
Therefore, in practice, it is generally hard to achieve the speedups suggested by roofline model (mentioned in \cref{tab:speedup_a100} and \cref{tab:speedup_nano} ) for sparse applications.

\section{Packed-Weight Sparse Convolution Kernel}
\label{sec:method}

The roofline analysis in Section~\ref{sec:roofline} shows that sparse $3\times3$
convolution sits at or below the memory bandwidth ridge point for both NVIDIA A100 as well as NVIDIA Tegra X1. Therefore, the bottleneck is not arithmetic but data movement. Realizing the theoretical FLOP and memory savings of sparsity therefore requires mitigating the two obstacles identified in Section~\ref{sec:intro} directly, rather than simply skipping zero-valued
multiplications in an otherwise dense kernel.

Our strategy combines three techniques, each targeting a specific source of
memory traffic or imbalance. 
\begin{itemize}
    \item \emph{Shared-memory tiling}: (Section~\ref{sec:tiling}) loads each region of the input feature map from DRAM exactly once per thread block and reuses it across all output channels and rows computed by that block, reducing redundant global memory traffic
    \item \emph{Packed weight format}: Section~\ref{sec:packed} fuses each nonzero's value and metadata into a single 32-bit word, so that a single coalesced memory transaction retrieves all the information a thread needs about a non-zero element.

    independent of sparsity level.
    \item \emph{Thread and warp mapping}: Strategy in (Section~\ref{sec:mapping}) distributes output channels across threads within a warp, mitigating obstacle (2), workload imbalance, when sparsity is comparable across channels. 
\end{itemize}

Together, these techniques aim to keep the kernel's realized memory traffic close to the lower bound implied by the roofline model, rather than the inflated traffic that naive sparse indexing would otherwise incur. Algorithm~\ref{alg:sparse-direct-conv} at the end of this section summarizes how these three techniques compose into the full kernel. Implementation details are explained in the following sections.
\begin{algorithm}[t]
\caption{Sparse Direct Convolution with Packed Weights}
\label{alg:sparse-direct-conv}
\begin{algorithmic}[1]
\Require Input tensor $X \in \mathbb{R}^{N \times IC \times H \times W}$
\Require Packed sparse weights $P$, offset array $O$
\Require Output dimensions $h,w$, number of output channels $OC$
\Ensure Output tensor $Y \in \mathbb{R}^{N \times OC \times h \times w}$

\ForAll{thread blocks $(b_x,b_y,b_n)$} \label{alg:inp}
    \State Determine batch index $n \gets b_n$ \label{alg:inp_}
    \State Determine output tile origin $(oh_0, ow_0)$ \label{alg:out_tile}
    \State Allocate shared-memory tile $S$ \label{ag:l4}

    \ForAll{threads in block}
        \State Cooperatively load required input tile from $X$ into $S$
        \State Use zero padding for out-of-bound input locations
    \EndFor
    \State Synchronize all threads \label{alg:l9}

    \ForAll{threads in block} \label{alg:l10}
        \State Determine warp id, lane id, output column $ow$
        \State Determine output-channel lane $oc_{\mathrm{lane}}$ \label{alg:l12}

        \For{$oc \gets oc_{\mathrm{lane}}$ to $OC-1$ step $32/COL\_PER\_WARP$} \label{alg:l13}
            \State Initialize accumulators $acc[0 \ldots R-1] \gets 0$

            \For{$j \gets O[oc]$ to $O[oc+1]-1$} \label{alg:l15}
                \State Unpack $P[j]$ into weight $v$, input channel $ic$, offsets $(p_x,p_y)$ \label{alg:l16}
                \State Compute shared-memory address for $S[ic, p_y, ow+p_x]$

                \For{$r \gets 0$ to $R-1$}
                    \State $acc[r] \gets acc[r] + v \cdot S[ic, p_y+r, ow+p_x]$
                \EndFor \label{alg:l20}
            \EndFor

            \For{$r \gets 0$ to $R-1$}
                \If{$oh_0+r < h$ and $ow < w$}
                    \State $Y[n, oc, oh_0+r, ow] \gets acc[r]$ \label{alg:l24}
                \EndIf
            \EndFor
        \EndFor
    \EndFor
\EndFor
\end{algorithmic}
\end{algorithm}



\subsection{Shared Memory Tiling}
\label{sec:tiling}
The input images are stored in $N \times IC \times H \times W$ layout to make this convolution compatible with major deep learning (DL) frameworks. Before computation, each block cooperatively loads the required input tile (a part of an image) into the shared memory. The tile includes the halo region required by the convolution window. For a kernel of size $K$, the shared memory tile is larger than the output tile by $K-1$ rows and columns if $stride=1$. This ensures that input tile is not loaded from DRAM multiple times by different threads. The data from the same tile is used for the entire computation block.

Input elements outside the feature map boundary are padded with zeros. In lines \ref{alg:inp} and \ref{alg:inp_} in \cref{alg:sparse-direct-conv}, we determine the corresponding image for each block and thread. The output tile indices ($oh_0, ow_0$) for a given thread are determined in line \ref{alg:out_tile} and then corresponding input tile is loaded by the entire block cooperatively in lines \ref{ag:l4} - \ref{alg:l9}.

\subsection{Thread and Block Organization}
\label{sec:mapping}
\begin{figure*}[t]
    \centering
\scalebox{1.0}{
    \input{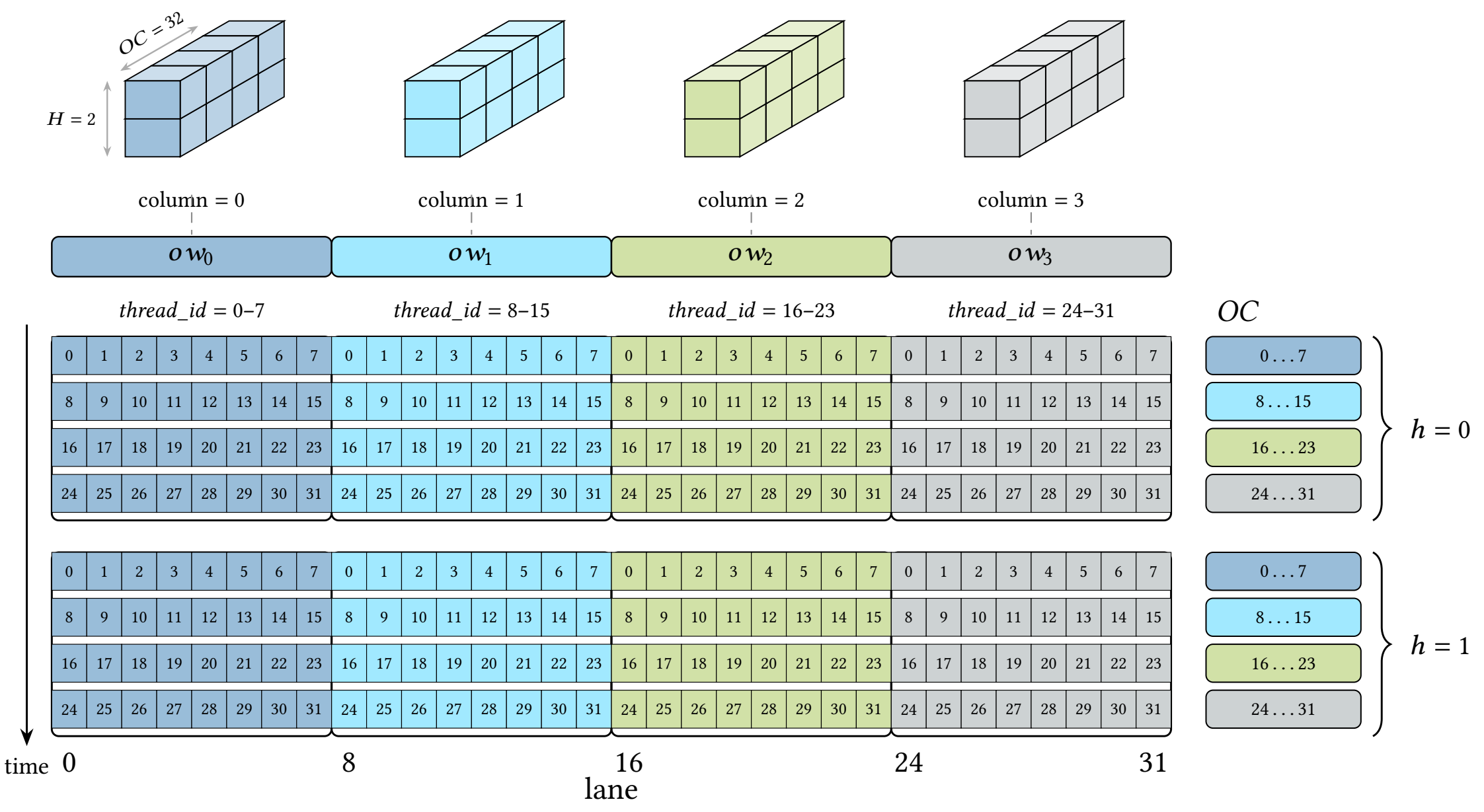}
}
\caption{
Mapping of an output tile to a warp. In this example, one warp computes four output columns across two output rows. The warp is partitioned into four groups of eight threads, with each group assigned to one output column. For the first column, threads $0$--$7$ initially compute output channels (OCs) $0$--$7$, before proceeding to OCs $8$--$15$, $16$--$23$, and $24$--$31$ in subsequent steps. Similarly, threads $8$--$15$, $16$--$23$, and $24$--$31$ process the second, third, and fourth output columns, respectively. The same mapping is repeated for every output row assigned to the warp. The number inside each thread box indicates the OC computed by that thread at the corresponding execution step.
}
    \label{fig:grid}
\end{figure*}

\begin{figure}[t]

    \centering
\scalebox{1.0}{
    \input{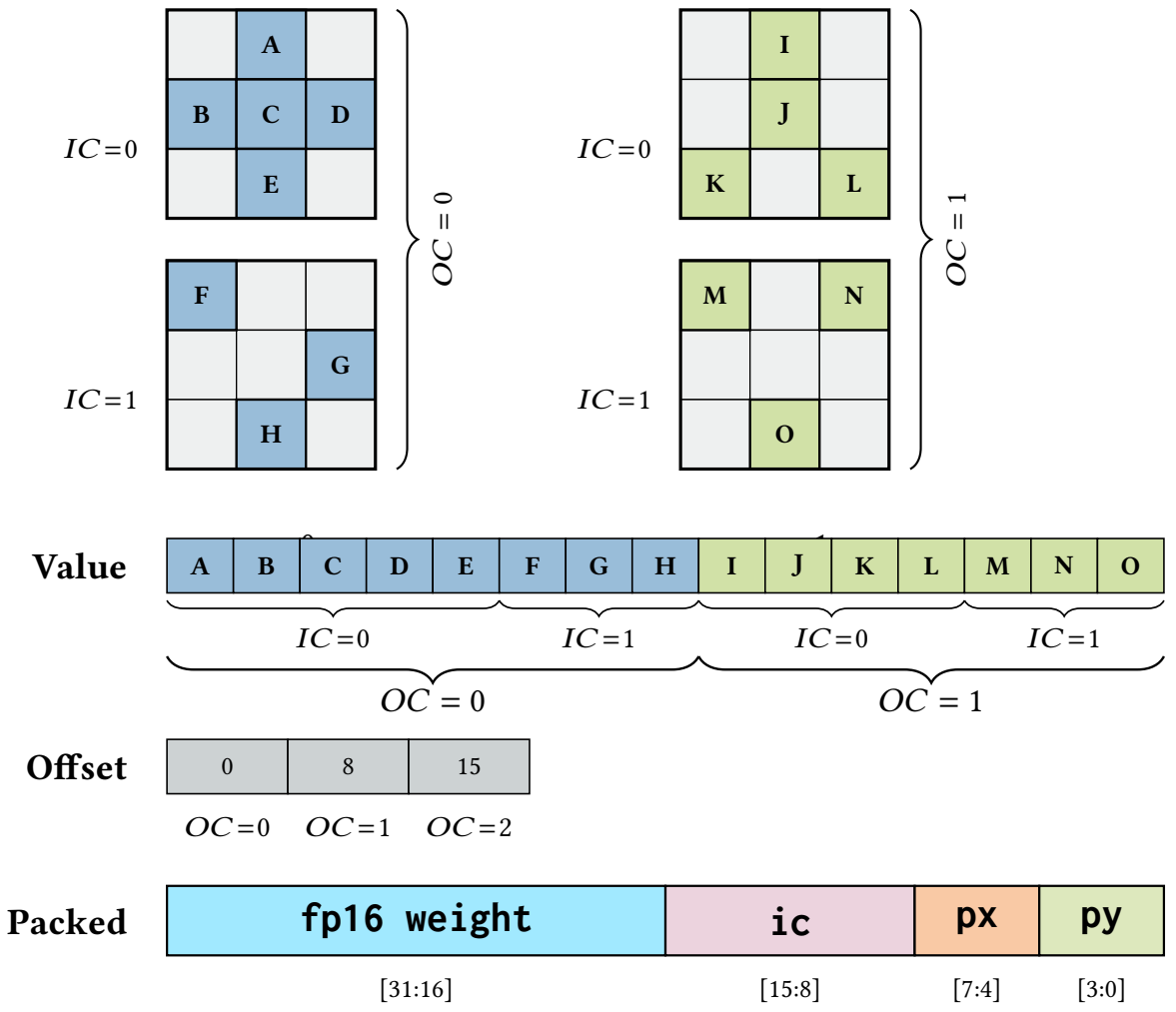}
}
    \caption{
 Each nonzero entry is encoded into a 32-bit packed word containing the FP16 weight value together with the input-channel index ($ic$) and kernel offsets ($px, py$). This layout enables aligned and coalesced memory accesses, allowing packed weights along with its indices to be fetched with a single 32-bit global memory transaction. Since sparse convolution is primarily memory bound, reducing memory traffic and improving load efficiency are critical for achieving high performance.
}
\label{fig:sp_weight_format} 
\end{figure}

The output feature map is divided into spatial tiles as shown in \cref{fig:grid}. The top grid shows how an output tile is mapped onto a warp. In this illustrative example, each warp computes four output rows and four output columns across all output channels. The actual mapping parameters were selected through an ablation study comparing multiple configurations. Although, we also parallelize over OCs along z-axis. Since a warp constitutes of 32 threads, eight threads collaborate to compute a single output column across multiple OCs and rows. Specifically, thread IDs 0–7 compute output channels 0–7 for the first column in the tile, and then proceed to compute output channels 8–15 (again for the first column in the tile), continuing in the same manner. Likewise, thread IDs 8–15 carry out the same computations for the second output column. This mapping strategy is applied consistently across all output tiles. Each thread block computes a tile containing several output columns and output rows for all output channels.
This mapping ensures that the computation is load-balanced when sparsity per output channel is similar. In lines \ref{alg:l10}-\ref{alg:l12} in the algorithm, one determines the corresponding output indices (row, col and output channel). In line \ref{alg:l13}, each thread loops over multiple OC with step $32/COL\_PER\_WARP$ as discussed above and as shown in \cref{fig:grid}

\subsection{Sparse Weight Format}
\label{sec:packed}
Each non-zero weight \emph{v}, along with its corresponding input channel \emph{ic}, \emph{px} and \emph{py} (respective x and y coordinates in the kernel) is stored as a single 32-bit unsigned integer. The upper 16 bits store the FP16 weight value, bits 15-8 store the input channel index, bits 7-4 store the kernel column offset, and bits 3-0 store the kernel row offset as shown in \cref{fig:sp_weight_format}. This is also shown in line \ref{alg:l16} of the pseudocode. Sparse convolution filters are transformed into a compact packed representation to improve memory efficiency, especially on resource-constrained devices.

This representation allows one 32-bit load to retrieve both the weight value and its metadata. An offset array of length $OC+1$ stores the start and end positions of the non-zero weights for each output channel, similar to the CSR format. All convolution arithmetic and accumulation are performed in FP32 using CUDA cores, and output values are stored in FP32. Upon loading a packed entry, the FP16 weight is converted to FP32 before computation.

\subsection{Sparse Accumulation}

Each thread traverses the non-zero weights belonging to its assigned output channels. For every packed entry, the thread unpacks the weight value, input channel index, and kernel offsets.

The corresponding input value is read from shared memory, and the result is accumulated using fused multiply-add (FMA) operations. FMA operations are especially helpful for larger GPUs such as NVIDIA A100 where AI is close to the ridge point as this reduces the number of operations per thread. Since the same weight is reused across multiple output rows, the method reduces repeated global memory accesses as shown in lines \ref{alg:l15}-\ref{alg:l20} of \cref{alg:inp}.

\subsection{Output Write-Back}

After accumulation, each thread writes the computed output values back to the output tensor as shown in line \ref{alg:l24}. Boundary checks are used for partial tiles near the image border. The output tensor is stored in $N \times OC \times H \times W$ layout as this is a widely used format and supported by major DL frameworks like PyTorch.





\section{Performance Results and Analysis}
\label{sec:results}

\begin{figure*}[!h]
    \centering
    \begin{subfigure}{0.48\textwidth}
        \includegraphics[width=\textwidth]{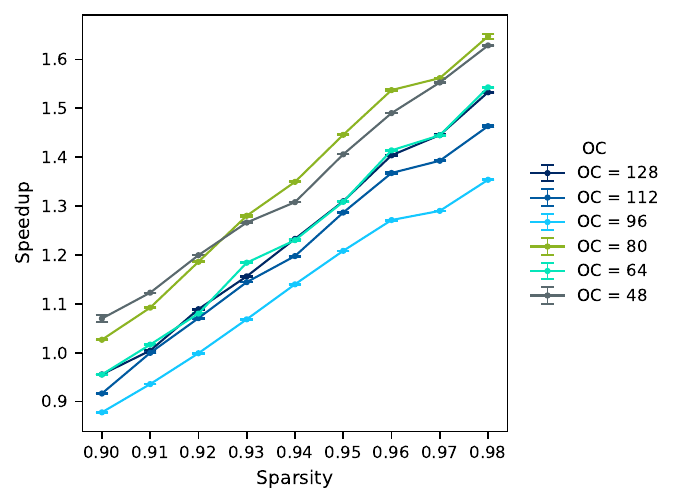}
        \caption{Speedup on NVIDIA A100}
        \label{fig:a100}
    \end{subfigure}
    \begin{subfigure}{0.48\textwidth}
        \includegraphics[width=\textwidth]{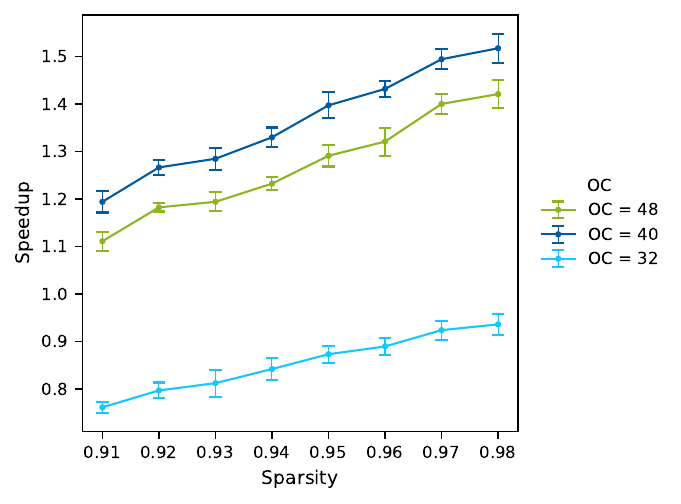}
        \caption{Speedup on Jetson Nano}
        \label{fig:tegra}
    \end{subfigure}
    \caption{Speedup relative to PyTorch for varying sparsities and Output Channels (OC) on NVIDIA A100 and Jetson Nano (NVIDIA Tegra X1) with configurations: $H=W=256$, $N=32$, $IC=32$ for NVIDIA A100 and $H=W=128$, $N=8$, $IC=16$ for Jetson Nano}
    \label{fig:bar}
\end{figure*}

\begin{figure*}[!t]
    \centering
    \begin{subfigure}{0.48\textwidth}
        \includegraphics[width=\textwidth]{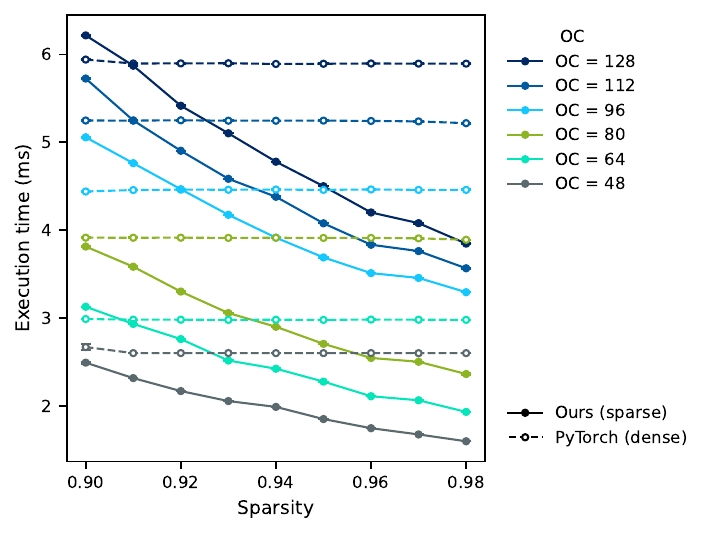}
        \caption{Execution times on NVIDIA A100}
        \label{fig:a100__exec}
    \end{subfigure}
    \begin{subfigure}{0.48\textwidth}
        \includegraphics[width=\textwidth]{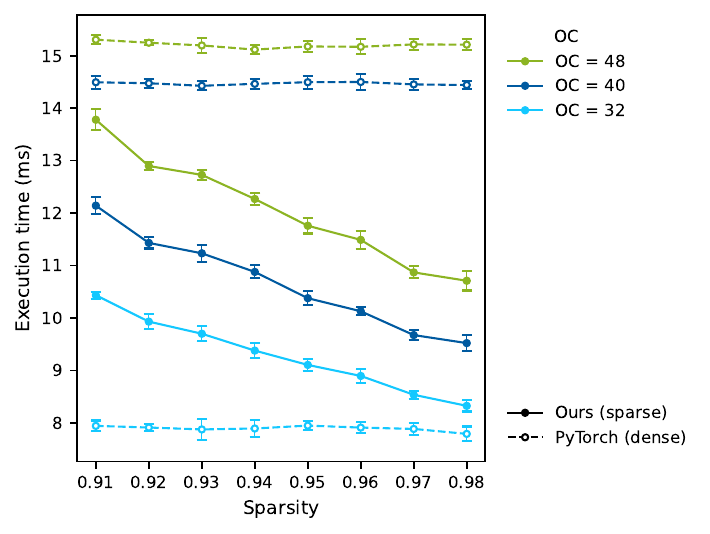}
        \caption{Execution times on Jetson Nano}
        \label{fig:tegra__exec}
    \end{subfigure}
    \caption{Execution times for varying sparsities and Output Channels (OC) on NVIDIA A100 and Jetson Nano (NVIDIA Tegra X1) with configurations: $H=W=256$, $N=32$, $IC=32$ for NVIDIA A100 and $H=W=128$, $N=8$, $IC=16$ for Jetson Nano}
    \label{fig:bar}
\end{figure*}

\begin{figure*}[!h]
    \centering
    \begin{subfigure}{0.48\textwidth}
        \includegraphics[width=\textwidth]{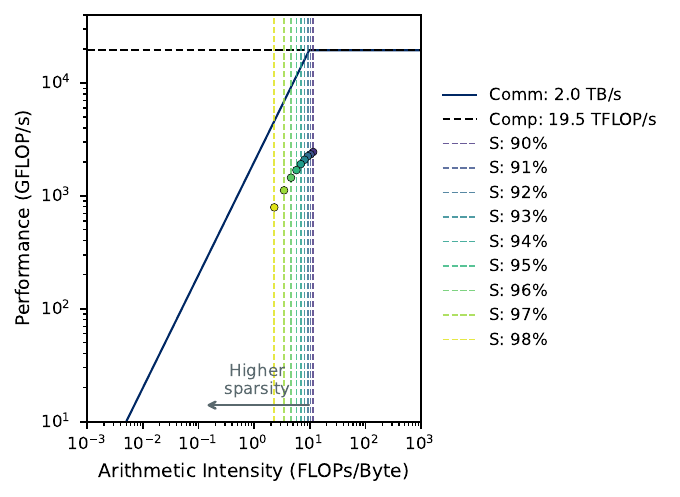}
        \caption{Performance on NVIDIA A100}
        \label{fig:a100_roof}
    \end{subfigure}
    \begin{subfigure}{0.48\textwidth}
        \includegraphics[width=\textwidth]{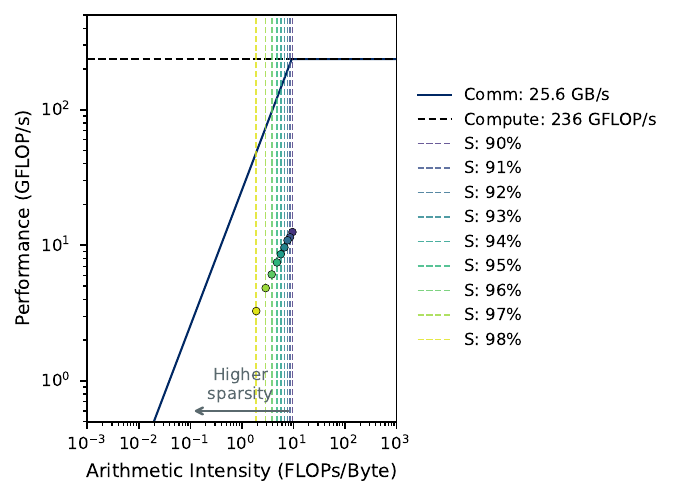}
        \caption{Performance on Jetson Nano}
        \label{fig:tegra_roof}
    \end{subfigure}
    \caption{Performance for varying sparsities on NVIDIA A100 and Jetson Nano (NVIDIA Tegra X1)}
    \label{fig:bar}
\end{figure*}

We benchmarked our sparse convolution implementation against PyTorch’s dense convolution baseline. To ensure a fair comparison, we developed Python bindings for our implementation using Pybind11~\cite{pybind11} and evaluated both approaches through the Python frontend. The convolution weights were generated using Python’s random module with varying sparsity levels. In this work, we report results for sparsity levels ranging from approximately 90\% to 98\% across different output channels (OC) and sparsities.

All benchmarks were conducted using CUDA cores. We evaluated our implementation on two NVIDIA platforms: the NVIDIA A100 GPU on the HoreKa supercomputer and the NVIDIA Jetson Nano (NVIDIA Tegra X1), which represents a more resource constrained edge device. For the NVIDIA A100 experiments, we used Python~3.11 together with CUDA~12.2, loaded via the environment module \texttt{devel/cuda/12.2} and PyTorch 2.9. The Jetson Nano experiments were conducted using JetPack 4.6, Python~3.6, and PyTorch~1.10. This is the latest software stack for this device. Rather than a limitation of our evaluation, we view this as representative of the deployment reality as it is fairly common for edge devices to be locked to whatever software stack is shipped with them. These software configurations were applied consistently during benchmarking.

In the following sections, we present benchmark results for both the NVIDIA A100 GPU and the NVIDIA Jetson Nano.

\subsection{NVIDIA A100}

Fig. \ref{fig:a100} shows the speedup achieved by the proposed sparse convolution implementation relative to PyTorch's dense convolution for varying sparsity levels and output-channel (OC) configurations on CUDA cores. The experiments were performed with fixed input dimensions of \[
H = W = 256,\qquad
N = 32,\qquad
IC = 32.
\]
where $H$ and $W$ are height and width of the input image, $N$ is the batch size and $IC$ is number of input channels
The results show that the performance benefit of the sparse implementation strongly depends on both the sparsity level and the number of output channels. For $OC=48$ and $OC=80$, the sparse implementation already outperforms the dense baseline at $90\%$ sparsity, achieving speedups of approximately $1.07\times$ and $1.03\times$, respectively. As sparsity increases, the speedup grows steadily, reaching $1.63\times$ for $OC=48$ and $1.65\times$ for $OC=80$ at $98\%$ sparsity.

For larger output-channel configurations, the sparse implementation becomes competitive at higher sparsity levels. At $90\%$ sparsity, the $OC=96$, $OC=112$, $OC=128$, and $OC=256$ configurations achieve speedups slightly below $1\times$. However, as sparsity increases, the sparse implementation gradually overtakes the dense baseline. The crossover point occurs near $91\%$ sparsity for $OC=64$ , $92\%$ for $OC=112$ and $OC=128$. Beyond these points, the performance advantage continues to increase with sparsity.

A dependence on the number of output channels can be observed. This behavior is expected since the current implementation parallelizes the OCs along the z-axis in the CUDA grid. The highest performance gains are achieved when the number of output channels can be processed within a single thread block, while larger output-channel configurations incur additional parallelization overhead. Nevertheless, the implementation consistently benefits from increasing sparsity and achieves speedups of up to approximately $1.65\times$ over PyTorch's dense convolution at $98\%$ sparsity while maintaining numerical correctness.

The error bars represent the standard error of the mean computed from ten independent benchmark runs, each consisting of forty convolution executions. The error bars are generally small, indicating stable and reproducible performance measurements across repeated executions.
Fig.~\ref{fig:a100__exec} compares the execution time of the proposed sparse convolution kernel with the dense PyTorch implementation on the NVIDIA A100. The runtime of the sparse convolution kernel decreases consistently as weight sparsity increases, while the dense baseline remains nearly constant since it performs the same computation regardless of sparsity. The improvement is most pronounced for larger output channel counts, demonstrating the ability of the proposed kernel to effectively exploit sparsity. This reduction in execution time directly translates into lower inference latency (time-to-solution).

Fig.~\ref{fig:a100_roof} presents the roofline analysis of the proposed sparse direct convolution kernel on the NVIDIA A100 GPU for OC=128. As weight sparsity increases from $90\%$ to $98\%$, the arithmetic intensity decreases from approximately 9.6 to 1.9 FLOP/Byte, shifting the kernel further toward the memory-bound region of the roofline. Correspondingly, the measured throughput decreases from approximately 2.4 TFLOP/s to 0.8 TFLOP/s, following the reduction in arithmetic intensity. The corresponding dense performance decreases from approximately 2.5 TFLOP/s to 0.5 TFLOP/s as, sparsity increases from $90\%$ to $98\%$ considering only non-zero computations. Nevertheless, the kernel remains well below both the theoretical HBM bandwidth roof and the FP32 compute roof, indicating that performance is constrained not only by raw memory bandwidth but also by irregular sparse memory accesses, reduced memory coalescing.

Overall, the results demonstrate that the proposed sparse direct convolution implementation effectively exploits high levels of weight sparsity on modern GPUs. 

\subsection{NVIDIA Jetson Nano}

Figure \ref{fig:tegra} shows the speedup achieved by the proposed sparse convolution implementation relative to PyTorch's dense convolution for varying sparsity levels and output-channel (OC) configurations on the NVIDIA Jetson Nano. The experiments were performed with fixed input dimensions of $H=W=128$, batch size $(N=8)$, and $IC=16$.

For Jetson Nano, we consider smaller dimensions since the memory on the device is limited. The results demonstrate that the performance benefit of the sparse implementation increases with sparsity for all evaluated OCs. Across the considered sparsity range of $91\% - 98\%$, the speedup grows steadily as the number of nonzero weights decreases. For the $OC=40$ configuration, the speedup increases from approximately $1.19\times$ at $91\%$ sparsity to $1.52\times$ at $98\%$ sparsity. Similarly, the $OC=48$ configuration improves from approximately $1.11\times$ to $1.42\times$, while the $OC=32$ configuration increases from 
$0.76\times$ to $0.94\times$.

A clear dependence on the number of output channels can also be observed. Configurations with larger output-channel counts achieve higher speedups than those with fewer output channels. This behavior is expected because larger output-channel counts increase the computational workload of the dense convolution, allowing the sparse implementation to benefit more substantially from the reduction in arithmetic operations and memory traffic.

The error bars represent the standard error of the mean computed from ten independent benchmark runs, each consisting of forty convolution executions. With the exception of a few isolated measurements, the error bars are relatively small, demonstrating stable and reproducible performance across repeated benchmark runs.

Fig.~\ref{fig:tegra__exec} compares the execution time of the proposed sparse convolution kernel with the dense PyTorch implementation on the Jetson Nano. The execution time of the proposed kernel decreases with increasing sparsity, whereas the dense PyTorch implementation remains largely unchanged. The sparse implementation consistently reduces the time-to-solution for inference across all evaluated output channel counts except $OC=32$.

Fig.~\ref{fig:tegra_roof} presents the roofline analysis of the proposed sparse direct convolution kernel on the Tegra X1 GPU for OC=48. As weight sparsity increases from $90\%$ to $98\%$, the measured throughput decreases from approximately 12.5 GFLOP/s to 3.3 GFLOP/s, following the reduction in arithmetic intensity. The corresponding dense performance decreases from approximately 11.5 GFLOP/s to 2.3 GFLOP/s as, sparsity increases from $90\%$ to $98\%$ considering only non-zero computations. Similar to the A100 case, the kernel remains well below both the theoretical HBM bandwidth roof and the FP32 compute roof, indicating that performance is constrained not only by raw memory bandwidth but also by irregular sparse memory accesses, reduced memory coalescing.

\begin{figure*}[!h]
    \centering
    \begin{subfigure}{0.48\textwidth}
        \includegraphics[width=\textwidth]{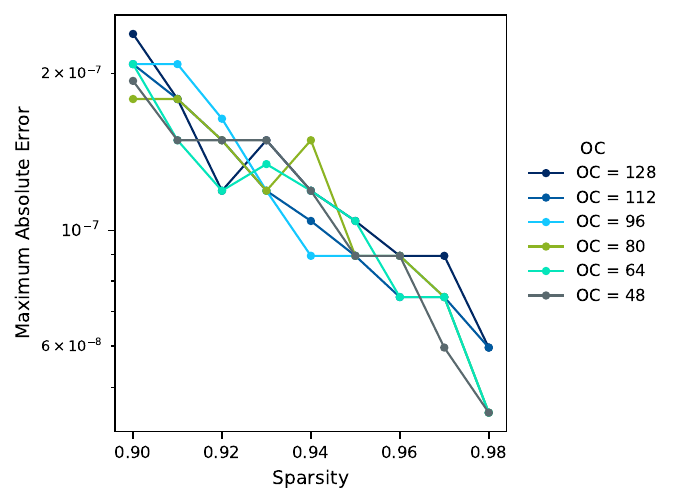}
        \caption{Maximum absolute error on NVIDIA A100}
        \label{fig:a100_err}
    \end{subfigure}
    \begin{subfigure}{0.48\textwidth}
        \includegraphics[width=\textwidth]{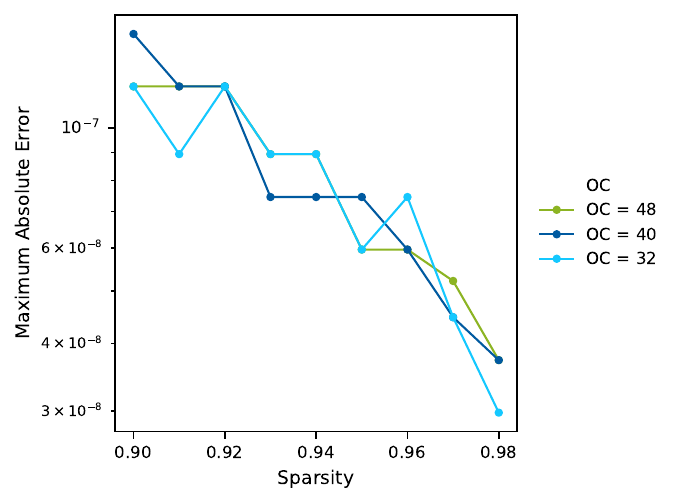}
        \caption{Maximum absolute error on Jetson Nano}
        \label{fig:tegra_err}
    \end{subfigure}
    \caption{Maximum absolute error for varying sparsities and Output Channels (OC) on NVIDIA A100 and Jetson Nano (NVIDIA Tegra X1) with configurations: $H=W=256$, $N=32$, $IC=32$ for NVIDIA A100 and $H=W=128$, $N=8$, $IC=16$ for Jetson Nano}
    \label{fig:bar}
\end{figure*}
\section{Conclusion}
\label{sec:conc}
This paper presented a theoretical and empirical analysis of sparse direct convolution for highly sparse convolutional neural networks on both: a high-performance GPU and an edge accelerator. Using roofline analysis, we demonstrated that increasing weight sparsity fundamentally changes the performance characteristics of convolution by reducing arithmetic intensity and shifting execution from the compute-bound to the memory-bound regime. Guided by this analysis, we proposed a direct sparse convolution kernel based on packed sparse weights, shared-memory tiling, and a warp-level thread mapping strategy. The resulting implementation achieved speedups of up to $1.65\times$ on the NVIDIA A100 and $1.52\times$ on the NVIDIA Jetson Nano compared to PyTorch's dense convolution implementation.

Beyond the observed kernel level speedups, our results highlight the practical potential of direct sparse convolution for edge AI applications. Modern convolutional neural networks such as ResNet-18, ResNet-34, and ResNet-50 are widely deployed in autonomous drones, mobile robots, medical devices, and intelligent sensors, where latency, power consumption, and limited computational resources are critical constraints. During inference, model weights remain resident in device memory while feature maps are streamed continuously. Consequently, reducing both the number of arithmetic operations and the amount of weight data transferred from memory can directly improve inference latency and throughput without modifying the network architecture. .

It is also important to note that these results are obtained for a $3 \times 3$ convolution kernel with a stride of $1$. Achieving similar performance improvements becomes increasingly challenging for larger input dimensions and kernel sizes. Although the roofline model predicts substantially larger speedups, the measured performance remains well below the theoretical memory-bandwidth roof. This gap reflects an inherent challenge of sparse computation rather than a limitation of the implementation itself. Roofline analysis assumes ideal memory behavior, perfect load balancing, and complete utilization of the available memory bandwidth. In practice, sparse workloads suffer from irregular memory accesses, reduced memory coalescing, lower cache locality, additional index decoding overhead, and workload imbalance caused by varying numbers of nonzeros across output channels. Moreover, NVIDIA GPUs and most modern edge accelerators are primarily optimized for regular dense computation. Efficiently exploiting high levels of sparsity may therefore require specialized hardware support, such as sparse processing units, sparsity-aware memory systems, and execution mechanisms capable of handling irregular workloads without sacrificing hardware utilization.

We also plan to evaluate complete sparse neural networks, including pruned ResNet models, to quantify end-to-end inference speedups, energy efficiency, and real-world deployment benefits on modern edge accelerators.

\begin{acks}  

This work was performed on the HoreKa supercomputer and the NHR@KIT Future Technologies Partition testbed, funded by the Ministry of Science, Research and the Arts Baden-Württemberg and by the Federal Ministry of Research, Technology and Space. This work is supported by the Helmholtz AI platform grant. The authors would like to thank the Federal Ministry of Education and Research and the state governments (www.nhr-verein.de/unsere-partner) for supporting this work/project as part of the joint funding of National High Performance Computing (NHR). The authors acknowledge the use of AI-based tools for language editing and grammar improvement during the preparation of this manuscript.
\end{acks}


\appendix
\section{Appendix}
\subsection{Maximum absolute error for convolution}
To validate the correctness of the convolution, \cref{fig:a100_err} and \cref{fig:tegra_err} report the maximum absolute error between the sparse convolution  kernel and the dense PyTorch baseline across the same sparsity sweep $H=W=256$, $N=32$, $IC=32$ for NVIDIA A100 and and $H=W=128$, $N=8$, $IC=16$ Jetson Nano respectively. Error decreases monotonically with sparsity for both of the devices. The residual error against the FP32 PyTorch baseline reflects two effects: the reduced number of accumulated terms at higher sparsity, and the one time quantization of weight values to FP16 during packing. Both of these factors diminish as sparsity increases, since fewer terms are summed.




\section*{AD Appendix}

\subsection*{Part 1: Overview of Contributions \& Artifacts}

\subsubsection*{A. Paper's Main Contributions}

The computational artifacts support the following contributions of the paper:

\begin{enumerate}
    \item A roofline analysis demonstrating that sparse direct convolution
    operates in the memory bandwidth bound region on modern GPUs. We showed the analysis for NVIDIA A100 and Jetso Nano (NVIDIA Tegra X1).

    \item A CUDA implementation of a sparse direct convolution kernel
    using
    \begin{itemize}
        \item packed 32-bit sparse weight representation,
        \item shared-memory tiling,
        \item an optimal thread mapping.
    \end{itemize}

    \item An experimental evaluation on NVIDIA A100 and NVIDIA Jetson Nano
    demonstrating execution time, speedup, numerical correctness, and
    roofline behavior compared to PyTorch's dense convolution.
\end{enumerate}

\subsubsection*{B. Computational Artifacts}

\begin{table}[t]
\centering
\caption{Overview of computational artifacts.}
\label{tab:artifacts}

\begin{tabularx}{\columnwidth}{p{2.7cm}cX}
\toprule
Artifact & Contr. & Paper elements reproduced \\
\midrule
Sparse convolution implementation &
2 &
Algorithm~1; Figures~3,4 \\

Roofline analysis scripts &
1 &
Figures~2,6; Tables~1,2,3 \\

Benchmark scripts &
3 &
Figures~5,6,7,8 \\

Plot generation scripts &
3 &
Performance figures and tables \\
\bottomrule
\end{tabularx}
\end{table}
\subsection*{Part 2: Artifact Identification}

\subsubsection*{Artifact 1: Sparse Convolution Evaluation}

\paragraph{A. Relation to Contributions}

The artifact reproduces the proposed sparse convolution kernel,
benchmarks it against PyTorch's dense convolution, performs roofline
analysis, and regenerates the figures presented in the paper.

\paragraph{B. Expected Results}

Executing the artifact should reproduce:

\begin{itemize}
    \item Increasing speedup of the sparse convolution kernel as sparsity increases.
    \item Execution times comparable to those reported in the paper.
    \item Numerical correctness compared with PyTorch.
    \item Roofline plots with similar performance trends.
\end{itemize}

Minor numerical differences may occur due to GPU architecture,
CUDA version, compiler optimizations, and runtime variability.

These results validate both the theoretical analysis and the proposed
sparse convolution implementation.

\paragraph{C. Expected Reproduction Time}

\begin{table}[h]
\centering
\caption{Estimated reproduction time.}
\begin{tabular}{lc}
\toprule
Step & Time (minutes) \\
\midrule
Artifact setup & 15--20 \\
Compilation & 3--5 \\
Benchmark execution & 2--15 \\
Result analysis & 2--3 \\
\midrule
Total & 40--70 \\
\bottomrule
\end{tabular}
\end{table}

\paragraph{D. Artifact Setup}

\textbf{Hardware}

\begin{itemize}
    \item NVIDIA GPU with CUDA support
    \begin{itemize}
        \item NVIDIA A100 (recommended)
        \item NVIDIA Jetson Nano (optional)
    \end{itemize}
\end{itemize}

\textbf{Software}

\begin{table}[h]
\centering
\caption{Required software for NVIDIA A100}
\begin{tabular}{ll}
\toprule
Software & Version \\
\midrule
CUDA Toolkit & 12.2 \\
Python & 3.11 \\
PyTorch & 2.9 \\
pybind11 & latest \\
CMake & $\ge$ 3.20 \\
GCC & $\ge$ 11 \\
\bottomrule
\end{tabular}
\end{table}

Repository:

The repository is hosted on GitHub at \href{https://github.com/keshvituteja/sparseconv}
{\textcolor{blue}{\url{https://github.com/keshvituteja/sparseconv}}}. The source code and plotting scripts used in the paper can be esily reproduced by following the using the scripts at \href{https://github.com/keshvituteja/sparseconv/tree/main/benchmarks/legacy}
{\textcolor{blue}{\url{https://github.com/keshvituteja/sparseconv/tree/main/benchmarks/legacy}}} and by following the instructions in \textit{legacy\_build.md}. The code can also be installed as a Python package, the detailed installation instructions are presented in \textit{README.md}.\\
\textbf{Datasets / Inputs}

No external datasets are required. Input tensors and sparse convolution kernels are generated automatically. Random sparse kernels are generated using a fixed random seed.



\textbf{Installation \& Deployment}

Requirements:

\begin{itemize}
    \item CUDA compiler (nvcc)
    \item CMake
    \item Python
    \item PyTorch
\end{itemize}

Compilation:


The Python bindings are compiled together with the CUDA implementation.






\bibliographystyle{unsrt}
\bibliography{main/references}










\end{document}